\documentclass[cits,a4paper]{PoS}
\usepackage{amsmath,amssymb}
\usepackage{graphicx}

\title{Computing the Adler function from the vacuum polarization function}

\ShortTitle{Computing the Adler function from the vacuum polarization function}

\author{\vspace{-0.75cm}\phantom{a} \hfill \textnormal{MITP/13-070}
  \newline \phantom{a} \hfill \textnormal{HIM-2013-06}
  \newline\newline \speaker{Hanno Horch}$^1$, Gregorio
  Herdo\'{i}za$^1$, Benjamin
  J\"ager$^{1,2}$, Hartmut Wittig$^{1,2}$\\
  $^{1}$PRISMA Cluster of Excellence, Institut f\"ur Kernphysik,
  Johannes
  Gutenberg Universit\"at Mainz, 55099 Mainz, Germany\\
  $^{2}$Helmholtz Institute Mainz, Johannes Gutenberg Universit\"at
  Mainz,
  55099 Mainz, Germany \\
  E-mail: \email{\{horch,herdoiza,jaeger,wittig\}@kph.uni-mainz.de}\\}

\author{Michele Della Morte\\
  CP3-Origins \& Danish IAS, University of Southern Denmark\\
  Campusvej 55, DK-5230 Odense M, Denmark and IFIC (CSIC)\\
  Calle Catedr\'{a}tico Jos\'{e} Beltr\'{a}n, 2. E-46980, Paterna, Spain\\
  E-mail: \email{dellamor@ific.uv.es}\\}

\author{Andreas J\"uttner\\
  School of Physics and Astronomy\\
  University of Southampton, UK\\
  E-mail: \email{a.juttner@soton.ac.uk}\\}

\abstract{We use a lattice determination of the hadronic vacuum
  polarization tensor to study the associated Ward identities and
  compute the Adler function.  The vacuum polarization tensor is
  computed from a combination of point-split and local vector
  currents, using two flavours of O($a$)-improved Wilson fermions.
  Partially twisted boundary conditions are employed to obtain a fine
  momentum resolution. The modifications of the Ward identities by
  lattice artifacts and by the use of twisted boundary conditions are
  monitored. We determine the Adler function from the derivative of
  the vacuum polarization function over a large region of momentum
  transfer $q^2$. As a first account of systematic effects, a
  continuum limit scaling analysis is performed in the large $q^2$
  regime.
  \newline
  \begin{flushright}\includegraphics[width=3cm]{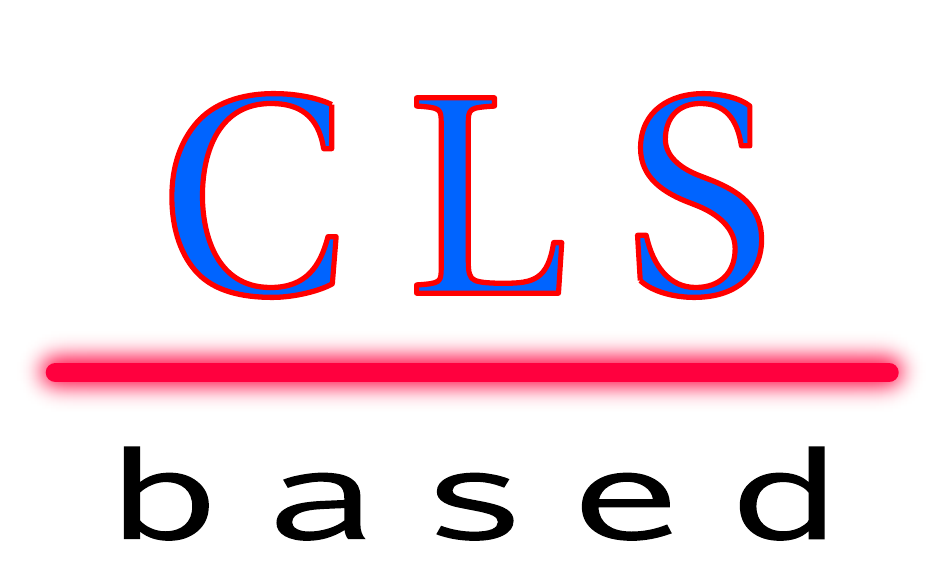}\end{flushright}}

\FullConference{31st International Symposium on Lattice Field Theory LATTICE 2013\\
                  July 29 - August 3, 2013\\
                  Mainz, Germany}

\begin{document}

\section{Introduction}
Recently, there has been a lot of interest in lattice determinations
of the hadronic vacuum polarization and related quantities, such as
the Adler function.  The latter can be used to determine the running
of $\alpha_{\textnormal{QED}}$~\cite{Adler:1974gd,Jegerlehner:2008rs},
which is a limiting factor for phenomenological studies at a future
linear collider. The Adler function is related to the vacuum
polarization function $\Pi(q^2)$ by
\begin{align}
  D(q^2)&=-\frac{3\pi q^2}{\alpha}\frac{d}{dq^2}\Delta
  \alpha^{\textnormal{had}}_{\textnormal{QED}}(q^2)=-12\pi^2q^2\frac{d}{dq^2}\Pi(q^2),
  \label{adlerfromvpf}
\end{align}
where $\Delta \alpha^{\textnormal{had}}_{\textnormal{QED}}(q^2)$ is
the shift of the fine structure constant due to hadronic
contributions. The present study reports on a continuation of the
project initiated in ref.~\cite{DellaMorte:2011aa}. \newline
In the continuum, the Ward identity for the vacuum polarization tensor
is given by
\begin{align}     
  &\sum_{\mu}q_\mu\Pi_{\mu\nu}(q^2)=\sum_{\nu}q_\nu\Pi_{\mu\nu}(q^2)=0.
  \label{wardidentityeq}
\end{align}     
On the lattice, these relations are not necessarily satisfied, for
instance, due to the use of a local non-conserved current and boundary
conditions. In this work we investigate the modifications of the Ward
identity due to the use of twisted boundary conditions.  The structure
of this work is as follows: in section~\ref{vpsection} we define the
basic quantities used in our study. We discuss the procedures
developed to determine the Adler function in
section~\ref{adlerfunctionsection}. We present our results for the
Ward identities in section~\ref{wisection}. In
section~\ref{conclusionssetion} we draw conclusions and give an
outlook for the future course of this project.

\section{The vacuum polarization}
\label{vpsection}
In our study we use two dynamical flavours of O$\left(a\right)$
improved Wilson fermions and the Wilson plaquette action. The
calculations are performed on gauge configurations generated by the
CLS initiative~\cite{clsinitiative}. The ensembles considered in this
paper are listed in table~\ref{clsconfigs}.

\begin{table}[ht!]
  \begin{center}
    \begin{tabular}{|c|c|c|c|c|c|c|}
      \hline
      Label  & $V/a^4$ &$\beta$ &   $a$\,[fm] & $m_\pi$\,[MeV] & $m_\pi L$ & $N_{\rm cfg}$\\
      \hline
      A5 & $64\times 32^3$&   5.20 & 0.079 & 312 & 4.0 & 250\\
      E5 & $64\times 32^3$&   5.30 & 0.063 & 451 & 4.7 & 168\\
      F6 & $96\times 48^3$&   5.30 & 0.063 & 324 & 5.0 & 217\\
      N6 & $96\times 48^3$&   5.50 & 0.050 & 340 & 4.0 & 173\\
      \hline
    \end{tabular}
    \caption{List of simulation parameters of the CLS ensembles
      considered in this work. The configurations were generated with
      $N_{\rm f}=2$, O$(a)$-improved Wilson fermions. The lattice
      spacing is taken from~\cite{Capitani:2011fg}.}
    \label{clsconfigs}
  \end{center}
\end{table}

The hadronic vacuum polarization tensor is defined as
\begin{align}
  \Pi^{N_{\rm f}}_{\mu\nu}(q^2)&=\int d^4xe^{iqx}\left<J^{N_{\rm
  f}}_\mu(x)J^{N_{\rm f}}_\nu(0)\right>=\left(g_{\mu\nu}q^2-q_\mu q_\nu\right)\Pi(q^2),
  \label{vpftensor}
\end{align}
with the vector currents $J^{N_{\rm f}}_\mu(x)=\sum_{f=1}^{N_{\rm
f}}Q_f\bar{\psi}_f(x)\gamma_\mu\psi_f(x)$, where $Q_f$ is the electric charge of each flavour. In the continuum, $\Pi(q^2)$ is
related to the vacuum polarization tensor via eq.~\eqref{vpftensor}, which
follows from Euclidean invariance and current conservation. When
eq.~\eqref{vpftensor} is evaluated on the lattice both connected and
disconnected diagrams occur. Despite the fact that the latter are
estimated~\cite{Francis:2013fzp,DellaMorte:2010aq} to be of the order of
$-10\%$, we currently neglect these contributions.
Following~\cite{DellaMorte:2011aa} we impose twisted boundary
conditions~\cite{Sachrajda:2004mi,Bedaque:2004ax,deDivitiis:2004kq} on
the quark fields
\begin{align}
  \psi(x_i+L)=e^{i\Theta_i}\psi(x_i)\quad\Rightarrow\quad
  \hat{q}_\mu=\frac{2}{a}\sin\left(\frac{\pi
      n_\mu}{L_\mu/a}-\frac{\Theta_\mu}{2L_\mu/a}\right),
  \label{tbc}
\end{align}
where the twist is only introduced in one direction
$\Theta=(0,\Theta_1, 0,0)$, to tune the momenta. The main benefit of
this is the improved constraint on fits in the small momentum region
between the first and second Fourier momentum. In the simulations, the
twist can be interpreted as a constant background field on the gauge
field, $U^{\Theta}_{\mu}(x)=U_\mu(x)e^{iaB_\mu}$, where $B_\mu$ is a
matrix in flavour space depending on the twist angles. In the case of
$N_{\rm f}=2$ we have $\psi^T(x)=\left(q^{(1)},q^{(2)}\right)^T$, thus
for our choice of twist angles we find $B_{\mu=0,2,3}=0$,
$B_1=\textnormal{diag}\left(B_1^{(1)},B_1^{(2)}\right)$ with
$B_1^{(j)}=\Theta^{(j)}_1/L$.\newline In the lattice regularization
there is a certain freedom for the implementation of these
currents. We use a combination of local and point-split currents
\begin{align}
  J^{(l),f}_\mu(x)&=Q_f\bar{\psi}_f(x)\gamma_\mu \psi_f(x),\\
  J^{(ps),f}_{\mu}(x)&=\frac{Q_f}{2}\left[\bar{\psi}_f(x+\mu)U^{\dagger}_\mu(x)e^{-iaB_\mu}(\gamma_\mu+1)\psi_f(x)+
    \bar{\psi}_f(x)U_\mu(x)e^{iaB_\mu}(\gamma_\mu-1)\psi_f(x+\mu)\right].\nonumber
  \label{pointspliteq}
\end{align}
The vacuum polarization tensor thus reads in our setup
\begin{align}
  \Pi^{(ps,l),N_{\rm f}}_{\mu\nu}(q^2)&=\int
  d^4xe^{iqx}\left<J^{(ps),N_{\rm f}}_\mu(x)J^{(l),N_{\rm
        f}}_\nu(0)\right>.
\end{align}
While the local current is not conserved, it allows us to reduce the
number of inversions needed for the determination of the vacuum
polarization, with respect to the case where the point-split current
would be used at both source and sink.

\section{The Adler function}
\label{adlerfunctionsection}
\begin{figure}
\input{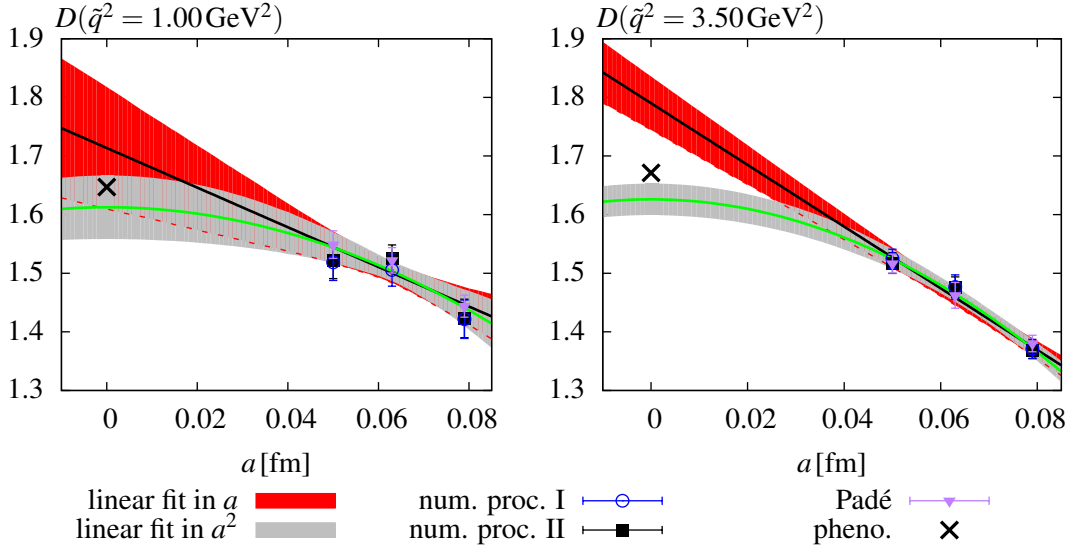}
\caption{Results for extrapolations to the continuum using linear fits
  in $a$ and in $a^2$ to the numerical procedure II at two different
  momentum transfers. For comparison the values determined by the
  other procedures for the Adler function are shown as well.}
\label{adlercontinuum}
\end{figure}

To determine the Adler function from eq.~\eqref{adlerfromvpf} it is
necessary to compute the derivative of the vacuum polarization
function. We have developed three different procedures to obtain the
derivative in order to check for systematic effects. For the first
procedure, we start by fitting a Pad\'{e}-Ansatz to the vacuum
polarization,
\begin{align}
  \Pi_{\rm fit}(q^2)&=c_0+q^2\left(\frac{c_1}{q^2+c_2^2}+\frac{c_3}{q^2+c_4^2}\right).
\end{align} 
For the other procedures, we profit from the fact that the use of
twisted boundary conditions yields a sufficient amount of data points
to determine the derivative of $\Pi(q^2)$ numerically in small steps
of $q^2$.
\begin{figure}
\input{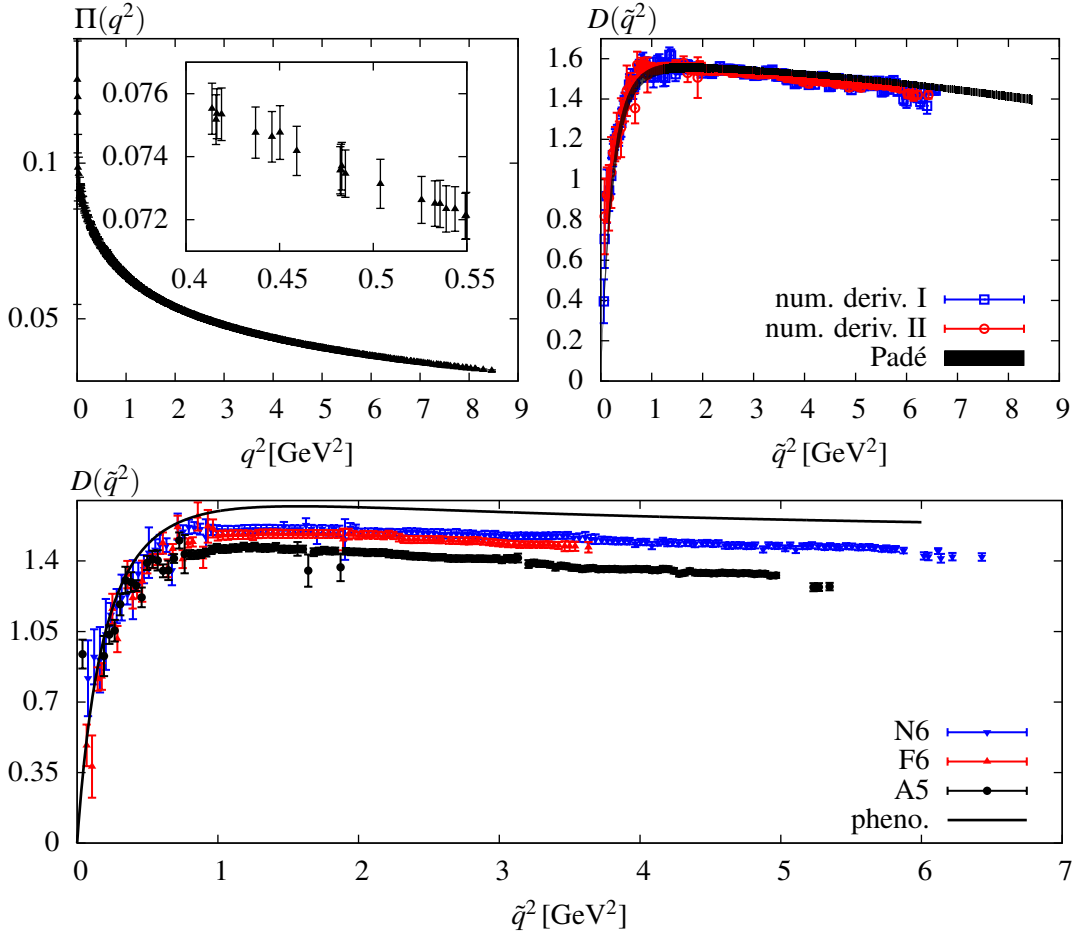}
\caption{Top left: Results for the vacuum polarization on the N6
  ensemble. Top right: Comparison of different procedures for the
  numerical derivative of $\Pi(q^2)$. Bottom: Results for numerical
  procedure II as an example for the three ensembles N6, F6, and A5
  with three different values of the lattice spacing,
  c.f.~table~\protect\ref{clsconfigs}. To compare the results to a
  phenomenological model~\cite{Francis:2013fzp} we rescale
  $q^2\rightarrow\tilde{q}^2=q^2(m^{\rm phys}_\rho/m^{\rm lat}_V)^2$,
  i.~e. the rho mass at the physical point divided by the vector meson
  mass measured individually on each ensemble.}
\label{alladlerplots}
\end{figure}
To compute the derivative we use fits at different values of $q^2$
separated by a certain step size $\epsilon$. At each value of $q^2$ we
use several fit intervals $q^2\pm\epsilon$, where
$\epsilon\in[0.02,1.0]\,$GeV$^2$.  In figure~\ref{alladlerplots} the
result for the vacuum polarization on the N6 ensemble is shown. We
stress that the fit window used in the small momentum region should be
small enough to describe the curvature of $\Pi(q^2)$, but not too
small so as to avoid strong fluctuations due to the limited number of
data points.  For the large momentum region we find that large fit
intervals are more suitable to describe $\Pi(q^2)$, because as $q^2$
is increased fewer points are available for the fit, and the curvature
is rather small. We use two different procedures to decide which fit
interval describes the numerical derivative best.\newline
The first numerical procedure uses linear fits,
$\Pi_{\rm fit}^{[l]}(q^2)=a_l+b_lq^2$. We look for a region in $\epsilon$
where the coefficient $b_l$ is stable.\newline
The second numerical procedure uses linear,
$\Pi_{\rm fit}^{[l]}(x)=a_l+b_lx$, and quadratic fits,
$\Pi_{\rm fit}^{[q]}(x)=a_q+b_qx+c_qx^2$, where
$x=\ln\left(q^2\right)$. The use of the variable $x$ is motivated by
the linear behaviour of $\Pi(x)$ in the interval of
$q^2\in[1,10]\,$GeV$^2$, and the second order term is used to choose
the appropriate fit window by constraining deviations from the linear
behaviour. In the top right panel of figure \ref{alladlerplots} we
compare the different procedures and find an overall good agreement
among the methods. The panel on the bottom of figure
\ref{alladlerplots} shows the result for the Adler function for three
different lattice spacings. Note that the ensembles considered in
figure \ref{alladlerplots} are not at a fixed pion mass as needed to
properly identify the lattice spacing dependence. The comparison to
the phenomenological curve should be regarded as qualitative at this
stage. In figure~\ref{adlercontinuum} the momentum transfer is
rescaled by $q^2\rightarrow\tilde{q}^2=q^2(m^{\rm phys}_\rho/m^{\rm lat}_V)^2$
\cite{Feng:2011zk} and the continuum extrapolation illustrated at two
different momentum transfers, $\tilde{q}^2=1.0\,$GeV$^2$ and
$\tilde{q}^2=3.5\,$GeV$^2$. As expected the signs of cut-off effects
increase with $q^2$. We apply linear fits in $a$ and in $a^2$ to test
the continuum limit scaling.

\section{The Ward identity of the vacuum polarization}
\label{wisection}
\begin{figure}
  \input{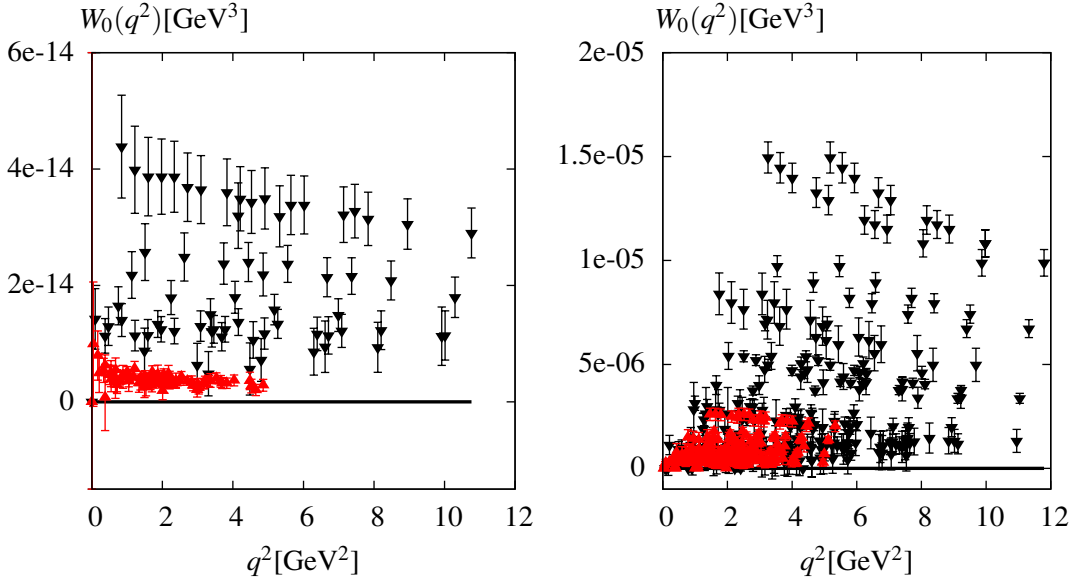}
  \caption{The determination of $W_0$ in~eq.~(\protect\ref{averagewi})
    for E5 with $L=2.0\,$fm are shown as red triangles facing up,
    while the black triangles facing down refer to results on F6 with
    $L=3.0\,$fm. Left: $W_0(q^2)$ at vanishing twist angle. Right:
    $W_0(q^2)$ for the largest twist angle of $\Theta=\frac{9\pi}{10}$
    that we use in simulations.}
  \label{wiscaling}
\end{figure}

\begin{figure}
  \input{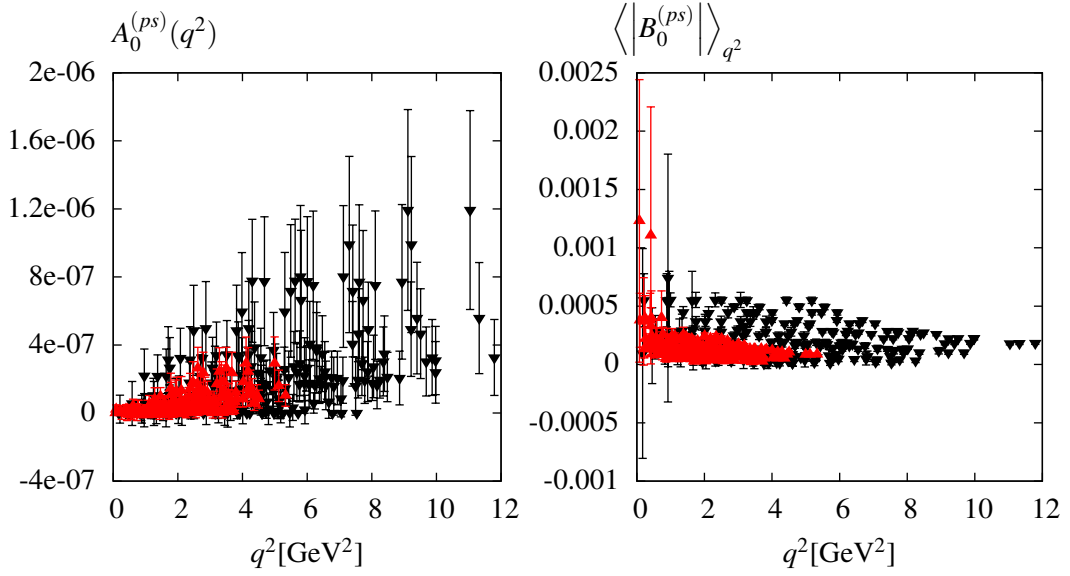}
  \caption{Results for $A^{(ps)}_\nu(q^2)$ and $B^{(ps)}_\nu(q^2)$ for
    $\Theta=\frac{9\pi}{10}$ are shown in the plots on the left and
    right, respectively. Black triangles facing up refer to results
    for E5, $L=2.0\,$fm, red triangles facing down to results for F6,
    $L=3.0\,$fm.}
  \label{wiratios}
\end{figure}

The introduction of twisted boundary conditions in the computation of
the vacuum polarization tensor requires the use of different twist
angles, $\Theta^{(1)}\neq\Theta^{(2)}$, in each of the two quark
propagators appearing in $\Pi_{\mu\nu}$. This leads to a breaking of
isospin symmetry which introduces modifications to the Ward identities
in eq.~\eqref{wardidentityeq}. The net effect of the twisted boundary
conditions on the Ward identity enters through the background field
$B_\mu^{(j)}=\Theta^{(j)}_\mu/L$. It should thus vanish in the
infinite volume limit. We perform a dedicated study of the lattice
Ward identity in order to monitor the possible impact on our
calculation of the vacuum polarization function. To this end we define
the quantity
\begin{align}
  W^{(ps)}_\nu(q^2)=\left<\left|\sum_{\mu}q_\mu\Pi^{(ps,l)}_{\mu\nu}(q^2)\right|\right>_{q^2},
  \label{averagewi}
\end{align}
where the absolute value of the sum is averaged over degenerate values
of $q^2$ to avoid compensating effects. The results shown in
figure~\ref{wiscaling} indicate that for vanishing twist angle the
Ward identity of the point-split current is fulfilled almost to
machine precision. For non-vanishing twist angles we observe that the
Ward identity in eq.~\eqref{wardidentityeq} is modified. We confirm
that this effect diminishes as the volume is increased.
To quantify the possible impact of the violation of the Ward identity
in the extraction of the vacuum polarization function $\Pi(q^2)$, we
consider the following dimensionless ratios
\begin{align}
  A^{(ps)}_\nu(q^2)&=\frac{W^{(ps)}_\nu(q^2)}{q_\nu
    q^2\left<\Pi(q^2)\right>_{q^2}},\quad
  B^{(ps)}_\nu(q^2)=\frac{\sum_{\mu}q_\mu\Pi^{(ps,l)}_{\mu\nu}}{q_\nu\Pi^{(ps,l)}_{\nu\nu}},
  \label{wiratioequations}
\end{align}
where the latter is similar to what was used
in~\cite{Aubin:2013daa}. Contrary to the case of $A^{(ps)}_\nu(q^2)$,
we observe that $B^{(ps)}_\nu(q^2)$ can lead to isolated peaks for
some values of $q^2$. This effect appears to be due to rather small
values of the denominator of $B^{(ps)}_\nu(q^2)$ in the case of
$\nu=1$ direction, where the twist is applied. We show the results for
these ratios in figure~\ref{wiratios}, and find that for the current
precision of our calculations, the violation of the Ward identity
induces a negligible effect on the determination of the vacuum
polarization function.

\section{Conclusions and outlook}
\label{conclusionssetion}
We presented three different methods to compute the Adler function
from vacuum polarization data which agree within errors over a large
range of momentum transfer. Furthermore we performed a preliminary
study of the continuum limit scaling in the large $q^2$ regime. In the
future we plan to extract the hadronic contribution to the running of
$\alpha_{\textnormal{QED}}$, which requires an extrapolation to the
continuum limit and a proper analysis of the
$m_\pi$-dependence.\newline
We presented numerical results for the Ward identities at different
lattice volumes at a single lattice spacing that signal modifications
of the usual Ward identities in the presence of twisted boundary
conditions. This effect diminishes as the volume is increased, and
given the current precision of our calculations of the vacuum
polarization it is observed to be negligible.\newline
Acknowledgements: Our calculations were performed on the ``Wilson''
HPC Cluster at the Institute for Nuclear Physics, University of
Mainz. We thank Dalibor Djukanovic and Christian Seiwerth for
technical support. We are grateful for computer time allocated to
project HMZ21 on the BlueGene computers \mbox{``JUGENE''} and
\mbox{``JUQUEEN''} at NIC, J\"ulich. This research has been supported
in part by the DFG via the SFB~1044.

\end{document}